\pdfoutput=1

\documentclass[12pt,a4paper]{article}

\usepackage{ifthen}
\newboolean{pdflatex}
\setboolean{pdflatex}{true}

\newboolean{articletitles}
\setboolean{articletitles}{true}

\newboolean{uprightparticles}
\setboolean{uprightparticles}{false}

\def\paperauthors{LHCb collaboration}
\def\paperasciititle{Test of lepton flavor universality with B -> K pi pi ell ell decays}
\def\papertitle{Test of lepton flavor universality with \decayLL decays}
\def\paperkeywords{{High Energy Physics}, {LHCb}}
\def\papercopyright{\the\year\ CERN for the benefit of the LHCb collaboration}
\def\paperlicence{CC BY 4.0 licence}
\def\paperlicenceurl{https://creativecommons.org/licenses/by/4.0/}

\newif\ifEnableSectionTOCLinks
\EnableSectionTOCLinksfalse

\usepackage[top=1in, bottom=1.25in, left=1in, right=1in]{geometry}

\usepackage{microtype}
\usepackage{lineno}  % for line numbering during review
\usepackage{xspace} % To avoid problems with missing or double spaces after
\usepackage{caption} %these three command get the figure and table captions automatically small

\usepackage{graphicx}  % to include figures (can also use other packages)
\usepackage{color}
\usepackage{colortbl}
\graphicspath{{./figs/}} % Make Latex search fig subdir for figures
\usepackage{amsmath} % Adds a large collection of math symbols
\usepackage{amssymb}
\usepackage{amsfonts}
\usepackage{upgreek} % Adds in support for greek letters in roman typeset

\newcommand*\patchAmsMathEnvironmentForLineno[1]{%
\expandafter\let\csname old#1\expandafter\endcsname\csname #1\endcsname
\expandafter\let\csname oldend#1\expandafter\endcsname\csname
end#1\endcsname
 \renewenvironment{#1}%
   {\linenomath\csname old#1\endcsname}%
   {\csname oldend#1\endcsname\endlinenomath}%
}
\newcommand*\patchBothAmsMathEnvironmentsForLineno[1]{%
  \patchAmsMathEnvironmentForLineno{#1}%
  \patchAmsMathEnvironmentForLineno{#1*}%
}
\AtBeginDocument{%
\patchBothAmsMathEnvironmentsForLineno{equation}%
\patchBothAmsMathEnvironmentsForLineno{align}%
\patchBothAmsMathEnvironmentsForLineno{flalign}%
\patchBothAmsMathEnvironmentsForLineno{alignat}%
\patchBothAmsMathEnvironmentsForLineno{gather}%
\patchBothAmsMathEnvironmentsForLineno{multline}%
\patchBothAmsMathEnvironmentsForLineno{eqnarray}%
}

\usepackage{hyperxmp}

\usepackage[pdftex,
            pdfauthor={\paperauthors},
            pdftitle={\paperasciititle},
            pdfkeywords={\paperkeywords},
            pdfcopyright={Copyright (C) \papercopyright},
            pdflicenseurl={\paperlicenceurl}]{hyperref}
\usepackage[colorinlistoftodos,textsize=scriptsize]{todonotes}

\usepackage[bottom,flushmargin,hang,multiple]{footmisc}

\usepackage[all]{hypcap} % Internal hyperlinks to floats.

\usepackage{xspace} 
\usepackage{upgreek}

\def\lhcb   {\mbox{LHCb}\xspace}

\def\MagUp {\mbox{\em Mag\kern -0.05em Up}\xspace}

\ifthenelse{\boolean{uprightparticles}}%
{

 \def\Pmu         {\ensuremath{\upmu}\xspace}                 
 \def\Pnu         {\ensuremath{\upnu}\xspace}                 
                  
 \def\Ppi         {\ensuremath{\uppi}\xspace}

 \def\Pchi        {\ensuremath{\upchi}\xspace}                 
 \def\Ppsi        {\ensuremath{\uppsi}\xspace}

 \def\PDelta      {\ensuremath{\Delta}\xspace}                 
 \def\PXi         {\ensuremath{\Xi}\xspace}                 
 \def\PLambda     {\ensuremath{\Lambda}\xspace}                 
 \def\PSigma      {\ensuremath{\Sigma}\xspace}                 
 \def\POmega      {\ensuremath{\Omega}\xspace}                 
 \def\PUpsilon    {\ensuremath{\Upsilon}\xspace}
 \let\oldPi\Pi
 \def\PPi         {\ensuremath{\oldPi}\xspace}

 \def\PB      {\ensuremath{\mathrm{B}}\xspace}                 
 \def\PD      {\ensuremath{\mathrm{D}}\xspace}                 
 \def\PJ      {\ensuremath{\mathrm{J}}\xspace}                 
 \def\PK      {\ensuremath{\mathrm{K}}\xspace}                 
 \def\PW      {\ensuremath{\mathrm{W}}\xspace}                 
 \def\PZ      {\ensuremath{\mathrm{Z}}\xspace}                 
 \def\Pb      {\ensuremath{\mathrm{b}}\xspace}

 \def\Pe      {\ensuremath{\mathrm{e}}\xspace}                 
 \def\Ps      {\ensuremath{\mathrm{s}}\xspace}

 \def\thebaroffset{0.0em}
}
{

 \def\Pmu         {\ensuremath{\mu}\xspace}                 
 \def\Pnu         {\ensuremath{\nu}\xspace}                 
                  
 \def\Ppi         {\ensuremath{\pi}\xspace}

 \def\Pchi        {\ensuremath{\chi}\xspace}                 
 \def\Ppsi        {\ensuremath{\psi}\xspace}                 
                  
 \mathchardef\PDelta="7101
 \mathchardef\PXi="7104
 \mathchardef\PLambda="7103
 \mathchardef\PSigma="7106
 \mathchardef\POmega="710A
 \mathchardef\PUpsilon="7107
 \mathchardef\PPi="7105
 \def\PB      {\ensuremath{B}\xspace}                 
 \def\PD      {\ensuremath{D}\xspace}                 
 \def\PJ      {\ensuremath{J}\xspace}                 
 \def\PK      {\ensuremath{K}\xspace}                 
 \def\PW      {\ensuremath{W}\xspace}                 
 \def\PZ      {\ensuremath{Z}\xspace}                 
 \def\Pb      {\ensuremath{b}\xspace}

 \def\Pe      {\ensuremath{e}\xspace}                 
 \def\Ps      {\ensuremath{s}\xspace}

 \def\thebaroffset{0.18em}
}
\newcommand{\offsetoverline}[2][\thebaroffset]{\kern #1\overline{\kern -#1 #2}}%

\makeatletter
\ifcase \@ptsize \relax% 10pt
  \newcommand{\miniscule}{\@setfontsize\miniscule{4}{5}}% \tiny: 5/6
\or% 11pt
  \newcommand{\miniscule}{\@setfontsize\miniscule{5}{6}}% \tiny: 6/7
\or% 12pt
  \newcommand{\miniscule}{\@setfontsize\miniscule{5}{6}}% \tiny: 6/7
\fi
\makeatother

\DeclareRobustCommand{\optbar}[1]{\shortstack{{\miniscule (\rule[.5ex]{1.25em}{.18mm})}
  \\ [-.7ex] $#1$}}

\def\en         {{\ensuremath{\Pe^-}}\xspace}   % electron negative (\em is taken)
\def\ep         {{\ensuremath{\Pe^+}}\xspace}

\def\mup        {{\ensuremath{\Pmu^+}}\xspace}
\def\mun        {{\ensuremath{\Pmu^-}}\xspace} % muon negative (\mum is taken)

\def\ellm       {{\ensuremath{\ell^-}}\xspace}
\def\ellp       {{\ensuremath{\ell^+}}\xspace}

\def\neu        {{\ensuremath{\Pnu}}\xspace}
\def\neub       {{\ensuremath{\overline{\Pnu}}}\xspace}

\def\neulb      {{\ensuremath{\neub_\ell}}\xspace}
\def\W      {{\ensuremath{\PW}}\xspace}

\def\Z      {{\ensuremath{\PZ}}\xspace}

\def\squark    {{\ensuremath{\Ps}}\xspace}

\def\bquark    {{\ensuremath{\Pb}}\xspace}

\def\pion   {{\ensuremath{\Ppi}}\xspace}
\def\piz    {{\ensuremath{\pion^0}}\xspace}
\def\pip    {{\ensuremath{\pion^+}}\xspace}
\def\pim    {{\ensuremath{\pion^-}}\xspace}

\def\kaon    {{\ensuremath{\PK}}\xspace}
\def\KorKbar {\kern \thebaroffset\optbar{\kern -\thebaroffset \PK}{}\xspace}

\def\Kp      {{\ensuremath{\kaon^+}}\xspace}
\def\Km      {{\ensuremath{\kaon^-}}\xspace}

\def\Dbar    {{\ensuremath{\offsetoverline{\PD}}}\xspace}
\def\D       {{\ensuremath{\PD}}\xspace}

\def\DorDbar {\kern \thebaroffset\optbar{\kern -\thebaroffset \PD}\xspace}

\def\Dzb     {{\ensuremath{\Dbar{}^0}}\xspace}
\def\Dp      {{\ensuremath{\D^+}}\xspace}
\def\Dm      {{\ensuremath{\D^-}}\xspace}

\def\DpDm    {\ensuremath{\Dp {\kern -0.16em \Dm}}\xspace}

\def\B       {{\ensuremath{\PB}}\xspace}

\def\BorBbar {\kern \thebaroffset\optbar{\kern -\thebaroffset \PB}\xspace}

\def\Bd      {{\ensuremath{\B^0}}\xspace}

\def\BdorBdbar {\kern \thebaroffset\optbar{\kern -\thebaroffset \Bd}\xspace}
\def\Bu      {{\ensuremath{\B^+}}\xspace}

\def\Bp      {{\ensuremath{\Bu}}\xspace}

\def\Bs      {{\ensuremath{\B^0_\squark}}\xspace}

\def\BsorBsbar {\kern \thebaroffset\optbar{\kern -\thebaroffset \Bs}\xspace}

\def\jpsi     {{\ensuremath{{\PJ\mskip -3mu/\mskip -2mu\Ppsi}}}\xspace}
\def\psitwos  {{\ensuremath{\Ppsi{(2S)}}}\xspace}

\def\Y#1S{\ensuremath{\PUpsilon{(#1S)}}\xspace}

\def\theX     {{\ensuremath{\Pchi_{c1}(3872)}}\xspace}

\def\LorLbar     {\kern \thebaroffset\optbar{\kern -\thebaroffset \PLambda}\xspace}

\newcommand{\decay}[2]{\mbox{\ensuremath{#1\!\to #2}}\xspace} 

\def\to                 {\ensuremath{\rightarrow}\xspace}

\def\qsq       {{\ensuremath{q^2}}\xspace}

\def\AT#1     {\ensuremath{A_{\mathrm{T}}^{#1}}\xspace}           % 2

\def\C#1      {\ensuremath{\mathcal{C}_{#1}}\xspace}                       % 9
\def\Cp#1     {\ensuremath{\mathcal{C}_{#1}^{'}}\xspace}                    % 7
\def\Ceff#1   {\ensuremath{\mathcal{C}_{#1}^{\mathrm{(eff)}}}\xspace}        % 9  
\def\Cpeff#1  {\ensuremath{\mathcal{C}_{#1}^{'\mathrm{(eff)}}}\xspace}       % 7
\def\Ope#1    {\ensuremath{\mathcal{O}_{#1}}\xspace}                       % 2
\def\Opep#1   {\ensuremath{\mathcal{O}_{#1}^{'}}\xspace}                    % 7

\newcommand{\aunit}[1]{\ensuremath{\text{\,#1}}}       
\newcommand{\tev}{\aunit{Te\kern -0.1em V}\xspace}
\newcommand{\gev}{\aunit{Ge\kern -0.1em V}\xspace}
\newcommand{\mev}{\aunit{Me\kern -0.1em V}\xspace}
\newcommand{\kev}{\aunit{ke\kern -0.1em V}\xspace}
\newcommand{\ev}{\aunit{e\kern -0.1em V}\xspace}
 
\newcommand{\mevc}{\ensuremath{\aunit{Me\kern -0.1em V\!/}c}\xspace}
\newcommand{\gevc}{\ensuremath{\aunit{Ge\kern -0.1em V\!/}c}\xspace}
\newcommand{\mevcc}{\ensuremath{\aunit{Me\kern -0.1em V\!/}c^2}\xspace}
\newcommand{\gevcc}{\ensuremath{\aunit{Ge\kern -0.1em V\!/}c^2}\xspace}
\newcommand{\gevgevcccc}{\ensuremath{\gev^2\!/c^4}\xspace} % for q^2

\def\fb   {\ensuremath{\aunit{fb}}\xspace}
\def\invfb   {\ensuremath{\fb^{-1}}\xspace}

\def\deriv {\ensuremath{\mathrm{d}}}

\def\gsim{{~\raise.15em\hbox{$>$}\kern-.85em
          \lower.35em\hbox{$\sim$}~}\xspace}
\def\lsim{{~\raise.15em\hbox{$<$}\kern-.85em
          \lower.35em\hbox{$\sim$}~}\xspace}

\def\pt         {\ensuremath{p_{\mathrm{T}}}\xspace}

\def\evtgen     {\mbox{\textsc{EvtGen}}\xspace}

\def\geant      {\mbox{\textsc{Geant4}}\xspace}

\def\photos     {\mbox{\textsc{Photos}}\xspace}

\def\pythia     {\mbox{\textsc{Pythia}}\xspace}

\def\tell1  {TELL1\xspace}
\def\ukl1   {UKL1\xspace}

\newcommand{\lhcborcid}[1]{\href{https://orcid.org/#1}{\hspace*{0.1em}\raisebox{-0.45ex}{\includegraphics[width=1em]{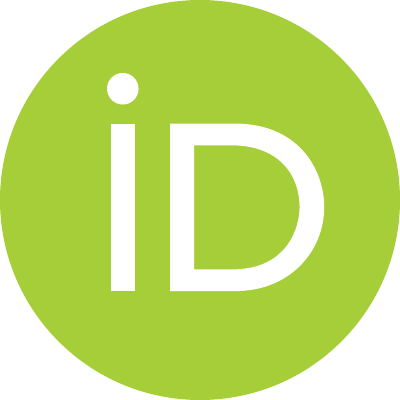}}}}

\newcommand{\BuKpipiGamma}{\decay{\Bu}{\Kp\pip\pim\gamma}\xspace}

\newcommand{\mkpi}{\ensuremath{m({\Kp\pim})}\xspace}
\newcommand{\mpipi}{\ensuremath{m({\pip\pim})}\xspace}
\newcommand{\mkpipi}{\ensuremath{m({\Kp\pip\pim})}\xspace}

\newcommand{\rkpipiinv}{{\ensuremath{R^{-1}_{\kaon\pion\pion}}}\xspace}

\def\rjpsi{{\ensuremath{r_{\PJ\mskip -3mu/\mskip -2mu\Ppsi\mskip 2mu}}}\xspace}

\def\Rpsitwos{{\ensuremath{R_{\psitwos}}}\xspace}

\newcommand{\decayEE}{\ensuremath{\Bu\to \Kp\pip\pim \ep \en }\xspace}
\newcommand{\decayMM}{\ensuremath{\Bu\to \Kp\pip\pim \mup \mun}\xspace}
\newcommand{\decayLL}{\ensuremath{\Bu\to \Kp\pip\pim \ell^+ \ell^-}\xspace}
\newcommand{\decayJpsiEE}{\ensuremath{\Bu\to \Kp\pip\pim\jpsi\left(\to \ep \en \right)}\xspace}
\newcommand{\decayJpsiMM}{\ensuremath{\Bu\to \Kp\pip\pim\jpsi\left(\to \mup \mun\right)}\xspace}
\newcommand{\decayJpsiLL}{\ensuremath{\Bu\to \Kp\pip\pim\jpsi\left(\to \ell^+ \ell^-\right)}\xspace}

\newcommand{\decaypsitwosLL}{\ensuremath{\Bu\to \Kp\pip\pim\psitwos\left(\to \ell^+ \ell^-\right)}\xspace}

\newcommand{\kpipi}{\ensuremath{\Kp\pip\pim}\xspace}
\hypersetup{
  colorlinks   = true, %Colours links instead of ugly boxes
  urlcolor     = blue, %Colour for external hyperlinks
  linkcolor    = blue, %Colour of internal links
  citecolor    = red   %Colour of citations
}

\ifEnableSectionTOCLinks
    \usepackage[explicit]{titlesec} % to change headings
    
    \let\oldcontentsline\contentsline
    \renewcommand

    \titleformat{\section}{\normalfont\Large\bf}{\hyperlink{tocsection.\thesection}{{\thesection} \parbox[t]{\dimexpr\textwidth-1pc}{#1}}}{1pc}{}

    \titleformat{\subsection}{\normalfont\bf}{\hyperlink{tocsubsection.\thesubsection}{{\thesubsection} \parbox[t]{\dimexpr\textwidth-1pc}{#1}}}{1pc}{}

    \titleformat{name=\section,numberless}[display]{}{}{0pt}{\normalfont\Huge\bfseries #1}
\fi

\usepackage{cite} % Allows for ranges in citations
\usepackage{mciteplus}
\usepackage{placeins}

\begin{document}

\renewcommand{\thefootnote}{\fnsymbol{footnote}}
\setcounter{footnote}{1}

% ===============================================================================
% Purpose: LHCb-PAPER journal paper title page template
% Author: 
% Created on: 2010-09-25
% ===============================================================================

%%%%%%%%%%%%%%%%%%%%%%%%%
%%%%%  TITLE PAGE  %%%%%%
%%%%%%%%%%%%%%%%%%%%%%%%%
\begin{titlepage}
\pagenumbering{roman}

% Header ---------------------------------------------------
\vspace*{-1.5cm}
\centerline{\large EUROPEAN ORGANIZATION FOR NUCLEAR RESEARCH (CERN)}
\vspace*{1.5cm}
\noindent
\begin{tabular*}{\linewidth}{lc@{\extracolsep{\fill}}r@{\extracolsep{0pt}}}
\ifthenelse{\boolean{pdflatex}}% Logo format choice
{\vspace*{-1.5cm}\mbox{\!\!\!\includegraphics[width=.14\textwidth]{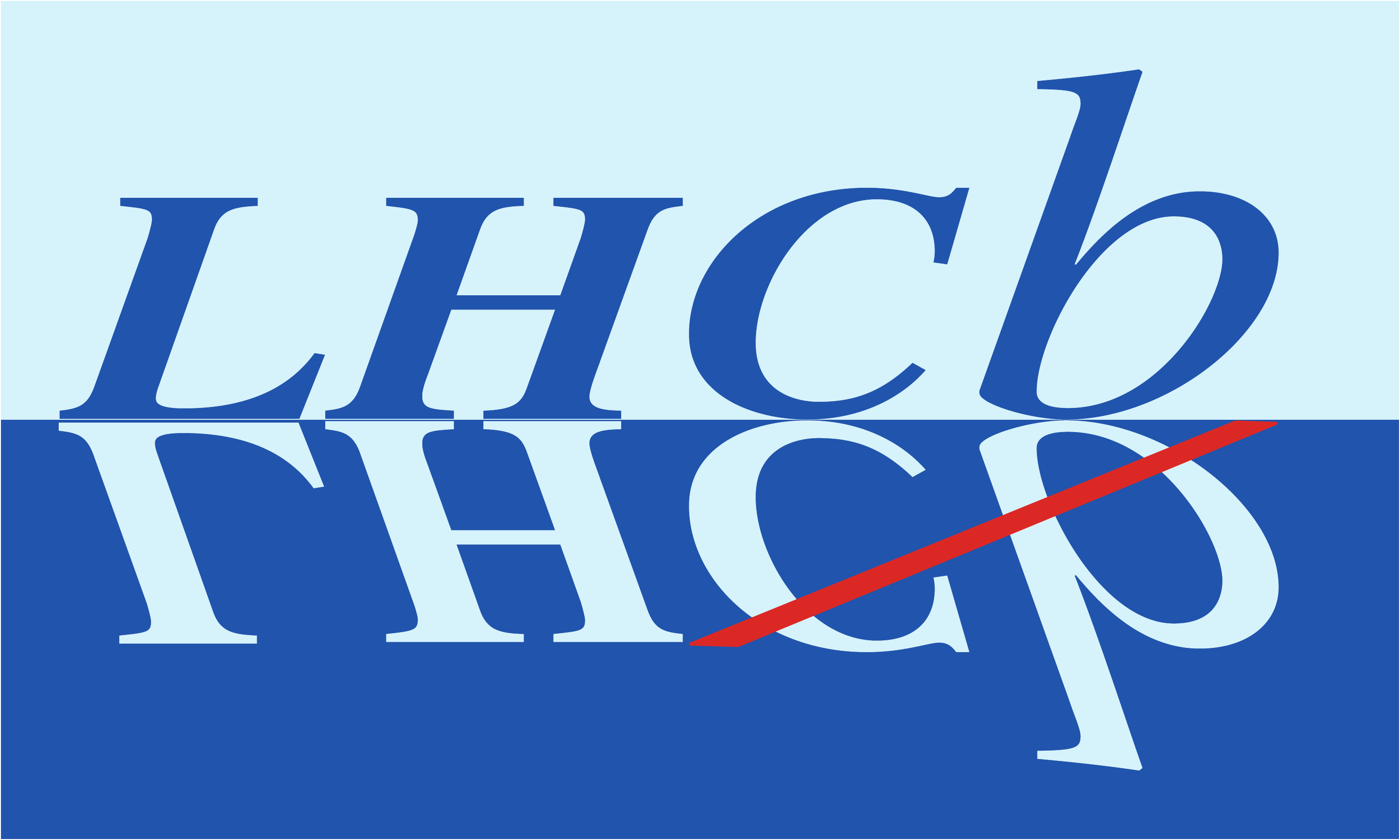}} & &}%
{\vspace*{-1.2cm}\mbox{\!\!\!\includegraphics[width=.12\textwidth]{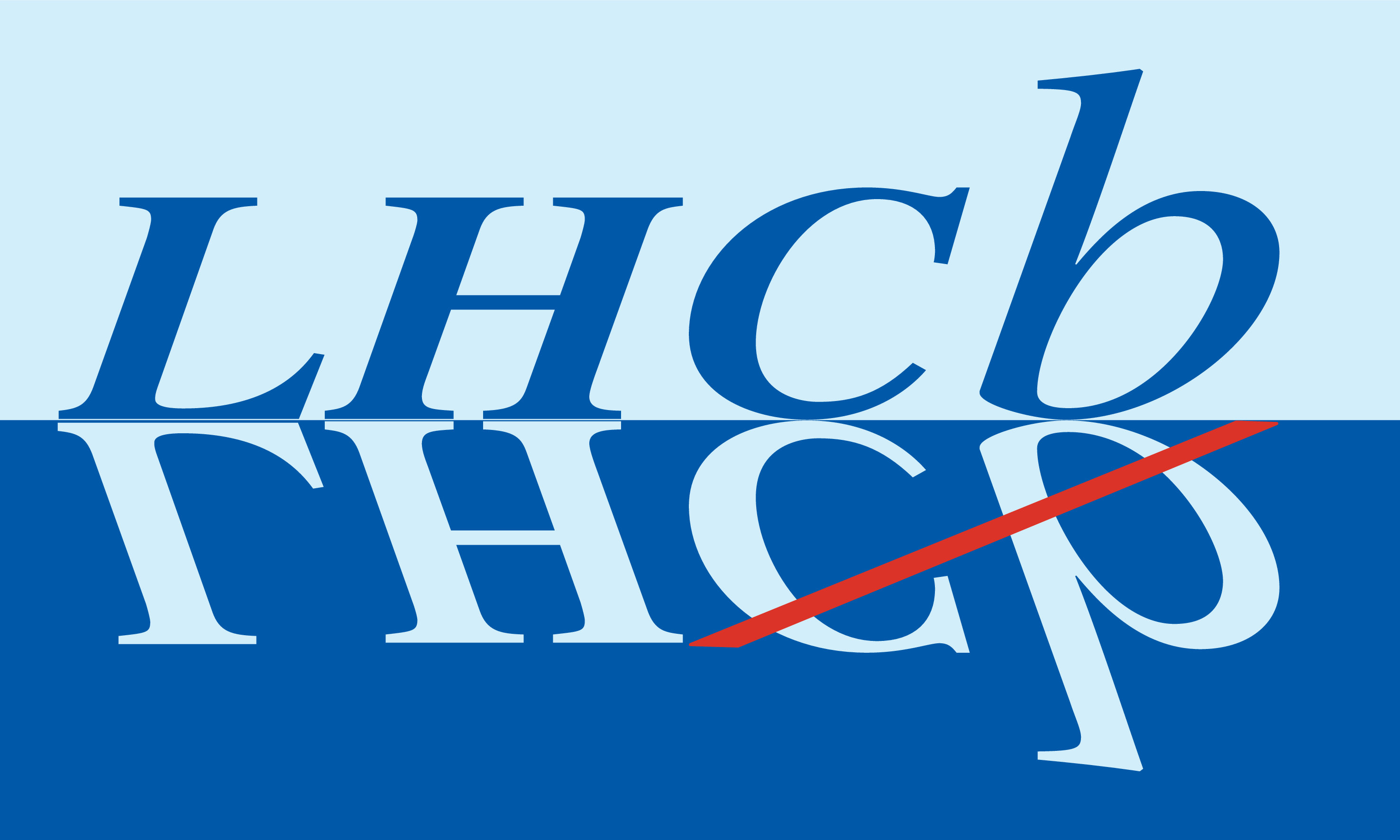}} & &}%
\\
 & & CERN-EP-2024-312 \\  % ID 
 & & LHCb-PAPER-2024-046 \\  % ID 
 & & May 12, 2025 \\ %Date of submission to Arxiv
 & & \\
% not in paper \hline
\end{tabular*}

\vspace*{4.0cm}

% Title --------------------------------------------------
{\normalfont\bfseries\boldmath\huge
\begin{center}
% DO NOT EDIT HERE. Instead edit macro in main.tex to keep metadata correct
  \papertitle 
\end{center}
}

\vspace*{2.0cm}

% Authors -------------------------------------------------
\begin{center}
%In the footnote, replace 'paper' by 'Letter' in case of submission to PRL or PLB 
% Edit macro in main.tex to keep metadata correct
\paperauthors\footnote{Full author list given at the end of the Letter.}
\end{center}

\vspace{\fill}

% Abstract -----------------------------------------------
\begin{abstract}
  \noindent
 The first test of lepton flavor universality between muons and electrons using \mbox{\decayLL} ($\ell=e,\mu$) decays is presented. The measurement is performed with data from proton-proton collisions collected by the \lhcb experiment at center-of-mass energies of 7, 8, and 13\tev, corresponding to an integrated luminosity of $9\invfb$.
 The ratio of branching fractions between \mbox{\decayEE} and \mbox{\decayMM} decays is measured in the dilepton invariant-mass-squared range \mbox{$1.1 < q^2 < 7.0\gevgevcccc{}$} and is found to be \mbox{$R_{K\pi\pi}^{-1} = 1.31^{+0.18}_{-0.17} \;(\mathrm{stat})\;^{+0.12}_{-0.09} \;(\mathrm{syst})$}, in agreement with the standard model prediction. The first observation of the \mbox{\decayEE} decay is also reported.
  
\end{abstract}

\vspace*{2.0cm}

\begin{center}
  Published in Physical Review Letters \textbf{134} (2025), 181803
  % JHEP /
  % Phys.~Rev.~D /
  % Phys.~Rev.~Lett. %/
%  Phys.~Lett.~B /
  % Eur.~Phys.~J.~C /
  %  Nucl.~Phys.~B /
  % Chin.~Phys.~C /
  % Nature~Physics /
  % sciPost~Physics /
  % J. Instr. /
  % Instruments 
\end{center}

\vspace{\fill}

{\footnotesize 
% Edit macro in main.tex to keep metadata correct
\centerline{\copyright~\papercopyright. \href{\paperlicenceurl}{\paperlicence}.}}
\vspace*{2mm}

\end{titlepage}

%%%%%%%%%%%%%%%%%%%%%%%%%%%%%%%%
%%%%%  EOD OF TITLE PAGE  %%%%%%
%%%%%%%%%%%%%%%%%%%%%%%%%%%%%%%%

%  empty page follows the title page ----
\newpage
\setcounter{page}{2}
\mbox{~}
%\newpage
%
%% Author List ----------------------------
%%  You need to get a new author list!
%\input{LHCb_authorlist.tex}
%
%The author list for journal publications is provided by the Membership Committee shortly after 'approval to go to paper' has been given.
%%It will be made available on the page
%%\verb!http://www.physik.uzh.ch/~strauman/forMemCo/LHCb-PAPER-XXXX-XXX/! .
%It will be sent to you by email shortly after a paper number has beens assigned.
%The author list should be included already at first circulation, 
%to allow new members of the collaboration to verify whether they have been included correctly.
%Occasionally a misspelled name is corrected or associated institutions become full members.
%In that case, a new author list will be sent to you.
%In case line numbering doesn't work well after including the authorlist, try moving the \verb!\bigskip! after the last author to a separate line.
%
%
%The authorship for Conference Reports should be ``The LHCb
%  collaboration'', with a footnote giving the name(s) of the contact
%  author(s), but without the full list of collaboration names.

\renewcommand{\thefootnote}{\arabic{footnote}}
\setcounter{footnote}{0}

\cleardoublepage

\pagestyle{plain}
\setcounter{page}{1}
\pagenumbering{arabic}

%% Comment before a final submission.
% \linenumbers

In the standard model (SM) of particle physics, flavor-changing transitions between quarks of the same charge cannot occur at leading order, but only through rare loop processes involving virtual weak bosons and quarks. As arbitrarily high virtual energies become accessible within loops, new physics~(NP) mediators could contribute to the decay process with comparable amplitudes to those of the SM, altering key physical observables from their expectations. Processes mediated by $b \to s \ell\ell$ quark-level transitions have drawn particular attention~\cite{PhysRevD.68.094016,PhysRevD.69.074020}, notably when testing lepton flavor universality (LFU), a key property of the SM. The SM electroweak couplings are predicted to be the same for processes including $e$, $\mu$, and $\tau$ leptons, after correcting for effects due to their different masses. Lepton flavor universality is an accidental symmetry of the SM, precisely verified up to $0.1\%$ in \W and \Z decays~\cite{W_LFU, Z_LFU}, kaon and pion decays~\cite{pi_LFU,K_LFU}, and in decays of charmonium resonances such as the \jpsi meson~\cite{Jpsi_LFU}. However, some scenarios beyond the SM predict a violation of LFU of up to $\sim 10\%$~\cite{LQ_16,Zprime_14,Zprime_15,Zprime_15_2,Zprime_15_3} in $b \to s$ transitions. Violation of LFU observed in ratios of decay rates for processes involving different lepton flavors, where uncertainties related to the hadronic form factors cancel out, would provide a model-independent signature of  NP in the decay.

Several ratios testing LFU with $b\to s\ell\ell$ transitions have been measured by the BABAR~\cite{PhysRevD.86.032012}, Belle~\cite{Belle_RK}, CMS~\cite{CMS_RK}, and LHCb collaborations~\cite{LHCb-PAPER-2014-024, LHCb-PAPER-2017-013, LHCb-PAPER-2019-009, LHCb-PAPER-2019-040, LHCb-PAPER-2021-004, LHCb-PAPER-2021-038, LHCb-PAPER-2022-045, LHCb-PAPER-2022-046, LHCb-PAPER-2024-032}.
In this Letter, the first LFU test using \mbox{\decayLL} decays is reported, where $\ell=e,\mu$. The inclusion of charge-conjugated processes is implied throughout. The \kpipi hadronic system, studied at Belle~\cite{Guler:2010if2} using \mbox{\decayJpsiLL} decays, is populated by several intermediate states such as the $\PK_1(1270)^+$ and $\PK_1(1400)^+$ resonances, which decay predominantly via the $K^*(892)^0\pip$ and $\Kp\rho(770)^0$ channels, thereby exhibiting a rich and complex spin structure. The present measurement, however, is inclusive as no resonant states are specifically identified in the hadronic system, which is selected in the \mbox{$1.1<\mkpipi< 2.4\gevcc$} kinematic region. Nevertheless, Ref.~\cite{Isidori_2022_inclusive} shows that the result can still be interpreted in the effective field theory framework, and it thus provides independent constraints on possible extensions of the SM. 

The measured observable is defined as
\begin{align}
    \rkpipiinv \equiv &
      \frac
          {\displaystyle\int _{1.1\,\gevgevcccc}^{7\,\gevgevcccc}
             \frac{\deriv\Gamma(\decayEE)}{\deriv\qsq}\deriv\qsq}
          {\displaystyle\int _{1.1\,\gevgevcccc}^{7\,\gevgevcccc}
             \frac{\deriv\Gamma(\decayMM)}{\deriv\qsq}\deriv\qsq } \;\;\;,
 \end{align}
where \qsq indicates the dilepton invariant-mass squared and $\Gamma$ the $q^2$-dependent decay rate.
The measurement of the inverse ratio \rkpipiinv is chosen to achieve better statistical properties given the low yield of the electron decay mode. To reduce the impact of systematic effects due to differences in the response of the LHCb detector to electrons and muons, the \rkpipiinv observable is in practice measured as a double ratio, by normalizing the signal-mode branching fractions to those of the corresponding $\jpsi$ control mode,
\begin{align}\label{eq:RKpipi}
R_{K\pi\pi}^{-1} & \equiv
\frac{\tfrac{\cal{N}}{\varepsilon}(\decayEE)}
     {\tfrac{\mathcal{N}}{\varepsilon}[\decayJpsiEE]}
\bigg{/}
\frac{\tfrac{\mathcal{N}}{\varepsilon}(\decayMM)}
     {\tfrac{\cal{N}}{\varepsilon}[\decayJpsiMM]}\, ,
\end{align}
where \mbox{$\tfrac{\cal{N}}{\varepsilon}$} are the efficiency-corrected yields. 
The measurement is performed using proton-proton ($pp$) collision data recorded by the \lhcb experiment between 2011 and 2018, corresponding to integrated luminosities of 1, 2, and 6\invfb at center-of-mass energies of 7, 8, and 13\tev, respectively. 
The decay \mbox{\decayMM} has been observed at LHCb with $3\invfb$ of data~\cite{LHCb-PAPER-2014-030}, while the first observation of the \mbox{\decayEE} mode is presented in this Letter.
In addition, the \mbox{\decayJpsiLL} and \mbox{\decaypsitwosLL} decays are used to calibrate the detector efficiencies and to validate the analysis procedure.

The LHCb detector is a single-arm forward spectrometer covering the pseudorapidity range $2 < \eta < 5$, described in detail in Refs.~\cite{LHCb-DP-2008-001,LHCb-DP-2014-002}. The online event selection is performed by a trigger, which consists of a hardware stage based on information from the calorimeter and muon systems, followed by a software stage which applies a full event reconstruction. At the hardware stage, events are required to have a muon with high \pt or a hadron, photon or electron with high transverse energy in the calorimeters. The software trigger requires a two-, three- or four-track secondary vertex with a significant displacement from any primary $pp$ interaction vertex. At least one charged particle must have a large transverse momentum and be inconsistent with originating from a primary vertex. A multivariate algorithm~\cite{BBDT,LHCb-PROC-2015-018} is used for the identification of secondary vertices consistent with the decay of a \bquark hadron.

Simulated data samples are used to evaluate the reconstruction and selection efficiencies entering the measurement. In addition, they are used to model the reconstructed-mass line shapes for signal candidates and to estimate possible background contributions. In the simulation, $pp$ collisions are generated using \pythia~\cite{Sjostrand:2007gs,*Sjostrand:2006za} with a specific \lhcb configuration~\cite{LHCb-PROC-2010-056}. Decays of unstable particles are described by \evtgen~\cite{Lange:2001uf}, in which final-state radiation is generated using \photos~\cite{davidson2015photos}. The interaction of the generated particles with the detector, and its response, are implemented using the \geant toolkit~\cite{Allison:2006ve, *Agostinelli:2002hh} as described in Ref.~\cite{LHCb-PROC-2011-006}. 

Simulated signal decays are generated with a uniform distribution in the phase space of the hadronic system.
To reproduce the resonant structure observed in data, the simulated samples are corrected using weights computed by training a gradient-boosted decision-tree reweighter~\cite{Rogozhnikov:2016bdp} on \mbox{\decayJpsiMM} decays. The correction uses the following quantities as inputs: the three invariant masses \mkpi, \mpipi and \mkpipi, the direction of the positively charged lepton in the $\ellp\ellm$ rest frame, the direction of the kaon in the $\Kp\pim$ rest frame, and the direction of the $\Kp\pim$ system in the \kpipi rest frame.
The simulated samples are further corrected to account for known discrepancies between simulation and data in the \Bp kinematics and event occupancy, as well as in the track reconstruction, particle identification and trigger efficiencies. In addition, the dilepton invariant-mass resolution is smeared to better match the data. The corrections are implemented in a similar fashion as in previous LFU measurements performed by the LHCb collaboration~\cite{LHCb-PAPER-2022-045}.

Candidate \mbox{\decayLL} decays are formed from combinations of five appropriately charged tracks that originate from a common vertex. The reconstructed \Bp candidate must be compatible with originating from a $pp$ collision when extrapolated according to its reconstructed momentum, and its decay vertex must be significantly displaced from that $pp$ collision vertex.
The final-state particles must satisfy a minimum $\pt$ requirement and be within the acceptance of the particle identification~(PID) system. Requirements on the PID-system response are also applied to discriminate between the different types of final-state particles. Electrons have a large probability to lose energy via bremsstrahlung. To mitigate such effects, reconstructed clusters in the calorimeter are added to the electron tracks if the cluster position is compatible with the extrapolated trajectory of the electron. Candidates are accepted if the reconstructed \Bp mass is in the range $[5150, 5600]\mevcc$ for the electron and muon control modes and the signal muon mode, and in $[4900, 5600]\mevcc$ for the signal electron mode.

To separate the signal from background composed of random combinations of tracks~(combinatorial), multivariate classifiers are trained separately for decay modes containing electrons or muons. The classifiers use vertex fit chi square, alongside kinematic and topological properties of the final-state particles and of the \Bp candidate. Simulated samples are used for modeling the signal, while to model the combinatorial background, the muon classifier is trained on selected candidates with an invariant mass greater than $5600\mevcc$. A similar approach is used for the electron classifier, but, due to the smaller training dataset, events in the $[4700, 4900]\mevcc$ range are also included.
To prevent any bias in the training of the classifiers, the $k$-fold approach is used~\cite{kfold}.
The working points of the classifiers are optimized to maximize the figure of merit $S/\sqrt{S+B}$, where $S$ and $B$ indicate the expected signal and background yields for a given classifier working point, respectively. As there is no {\it a priori} branching fraction measurement for the electron signal mode, its expected yield is instead inferred from the muon signal branching fraction assuming LFU. 

The \qsq range used to select signal decays extends up to 7\gevgevcccc, above the \qsq boundary used in similar LFU tests~\cite{LHCb-PAPER-2014-024, LHCb-PAPER-2017-013, LHCb-PAPER-2019-009, LHCb-PAPER-2019-040, LHCb-PAPER-2021-004, LHCb-PAPER-2021-038, LHCb-PAPER-2022-045, LHCb-PAPER-2022-046, LHCb-PAPER-2024-032}, in order to maximize the yield of the electron mode. However, this choice leads to an increase in the number of \mbox{\decayJpsiLL} decays migrating into the signal region.
To reduce such leakage, the invariant mass of the \Bp candidates, reconstructed by constraining the dilepton mass to the known \jpsi mass value, is required to lie outside the [5139.3, 5476.2]\mevcc range, which contains 95\% of simulated \mbox{\decayJpsiEE} events.

Background arising from fully hadronic \Bp decays proceeding through an intermediate \D meson, such as \mbox{$\Bp\to\Dzb(\to \Kp\pim)\pip\pim\pip$} or \mbox{$\Bp\to \Dm(\to \Kp\pim\pim)\pip\pip$} decays, where two pions are wrongly identified as leptons, is reduced by removing candidates where the invariant mass of the intermediate \D meson, calculated by assigning the pion mass to the leptons, falls within $\pm 50$\mevcc of the known $D$ meson mass~\cite{PDG2024}. Similarly, candidates arising from a swapped lepton and hadron of the same charge in \mbox{\decayJpsiLL} decays are removed by applying a veto on the lepton-hadron invariant mass, where the lepton mass is assigned either to the kaon or pion. Background from \mbox{$\Bp\to \phi(\to \Kp\Km) \Kp\ellp\ellm$} decays where two kaons are misidentified as pions is found to be negligible.
Semileptonic decays such as \mbox{$\Bp\to \Dm(\to \Kp\pim\ellp\neulb)\pip\ellp\neu$} are found to already be efficiently rejected by the selection requirements. Additional requirements are implemented when selecting control \mbox{\decayJpsiLL} decays, to reject backgrounds from \mbox{$\Bp\to \Kp\psitwos[\to\pip\pim\jpsi(\to\ellp\ellm)]$} and \mbox{$\Bp\to \Kp \theX[\to\pip\pim\jpsi(\to\ellp\ellm)]$} transitions. 
Less than 1\% of events satisfying all selection criteria have multiple \Bp candidates. In such cases, only one candidate is retained, chosen arbitrarily. The efficiency of the above selection is observed to increase with \qsq.

Signal and control yields are extracted from unbinned maximum-likelihood fits to the invariant-mass spectrum of \Bp candidates passing the full selection. The invariant-mass resolution is improved for the control modes by constraining the dilepton mass to that of the relevant \jpsi or \psitwos meson.

The analysis procedure is first validated by measuring \rjpsi, the ratio of the \mbox{\decayJpsiMM} to \mbox{\decayJpsiEE} branching fractions, and \Rpsitwos, the double ratio with the \mbox{\decayJpsiLL} and \mbox{\decaypsitwosLL} branching fractions, constructed in the same way as Eq.~(\ref{eq:RKpipi}), but with \mbox{\decaypsitwosLL} decays instead of \mbox{\decayLL} decays.

The ratio of yields of the two channels  \mbox{\decayJpsiLL} is determined with a simultaneous fit to the invariant-mass spectra of \mbox{\decayJpsiEE} and \mbox{\decayJpsiMM} candidates, selected with $q^2 \in[7,11]\gevgevcccc$. For these decays, the candidate-\Bp invariant-mass spectrum is modeled using a linear combination of two crystal ball (CB) functions~\cite{Skwarnicki:1986xj}. With the core mean and resolution shared between CB functions, all line shape parameters are fixed from fits to simulated samples.
However, to take into account data-simulation differences, a mean shift parameter and resolution scale factor are introduced for the fit to data. The combinatorial background component is modeled with an exponential function, with the slope parameter free to vary.  After correcting for the efficiency, the \rjpsi ratio is found to be $\rjpsi = 1.033 \pm 0.017$, where the quoted uncertainty is statistical. No significant variation of \rjpsi is observed in regions (bins) of all relevant kinematical and geometrical variables, with the \rjpsi values found to be compatible with unity for all variables considered. Deviations from unity among the bins are retained as a systematic uncertainty on \rkpipiinv.

The double ratio $\Rpsitwos$, is measured from an additional simultaneous fit using the pairs of decay modes \mbox{\decayJpsiLL} and \mbox{\decaypsitwosLL}, where the latter is selected with $\qsq\in[11,15]\gevgevcccc$. The invariant-mass line shape for the \mbox{\decaypsitwosLL} modes is modeled in the same way as described for the \mbox{\decayJpsiLL} modes. The efficiency-corrected double ratio is measured to be  $\Rpsitwos = 1.040 \pm 0.030$, where the quoted uncertainty is statistical. The agreement of the \rjpsi and \Rpsitwos ratios with unity is thus consistent with the expectations of LFU in charmonium decays.

The \rkpipiinv observable is determined from a simultaneous fit to the invariant-mass spectra of signal \mbox{\decayLL} and control \mbox{\decayJpsiLL} decays. The model describing the control component is also adopted for the signal modes, with shape parameters fixed from fits to simulated signal decays. To account for simulation-data differences a mean shift and a resolution scale factor are included. These two parameters are fixed to the values found in \mbox{\decayJpsiLL} decays,  where the \jpsi mass constraint on the dilepton system is removed as its impact on the resolution would be inapplicable to signal. The combinatorial background is again described by an exponential function, whose slope is allowed to vary. While no other contributions are expected in the mass spectrum of \mbox{\decayMM} decays, the lower yield and worse resolution of the \mbox{\decayEE} mode necessitate a larger fit region, leading to the presence of additional sources of background. These include leakage from \mbox{\decayJpsiEE} events into the signal \qsq region; background from partially reconstructed \Bp decays, in which one or more final-state particles are not reconstructed; and residual background due to hadrons misidentified as electrons. 

The mass line shape of the leakage component is modeled using simulated samples with a kernel-density estimator~\cite{Cranmer_2001}, and kept fixed in the fit. Its normalization is fixed to the yield observed in the \mbox{\decayJpsiEE} mode, scaled for the efficiency of the signal selection applied to the control mode. 

The partially reconstructed background line shape is modeled by the convolution of an Argus function~\cite{ALBRECHT1990278} with a Gaussian function.
This mass line shape is fixed from a fit to data selected in the $\qsq\in[7,11]~\gevgevcccc$ region,
where the control \mbox{\decayJpsiEE} decay mode is explicitly removed by requiring the reconstructed \Bp mass, with the dielectron mass constrained to the \jpsi meson mass, to be smaller than $5150\mevcc$. The obtained sample is enriched in partially reconstructed decays as a result. The contribution from higher-mass charmonium states, such as \mbox{$\Bp\to \psitwos(\to \jpsi\pip\pim) \PK^*(892)^+ (\to \Kp\piz)$}, is then subtracted, since such feeddown decays are not present for the signal decay mode.
The fit is repeated without vetoing \mbox{\decayJpsiEE} decays to extract the ratio of the remaining partially reconstructed background with respect to the control yield. This ratio ultimately constrains the relative amount of partially reconstructed background with respect to the signal yield in the fit to the signal decay mode.
\begin{figure}[!t]
   \begin{center}
     \includegraphics[width=0.45\linewidth]{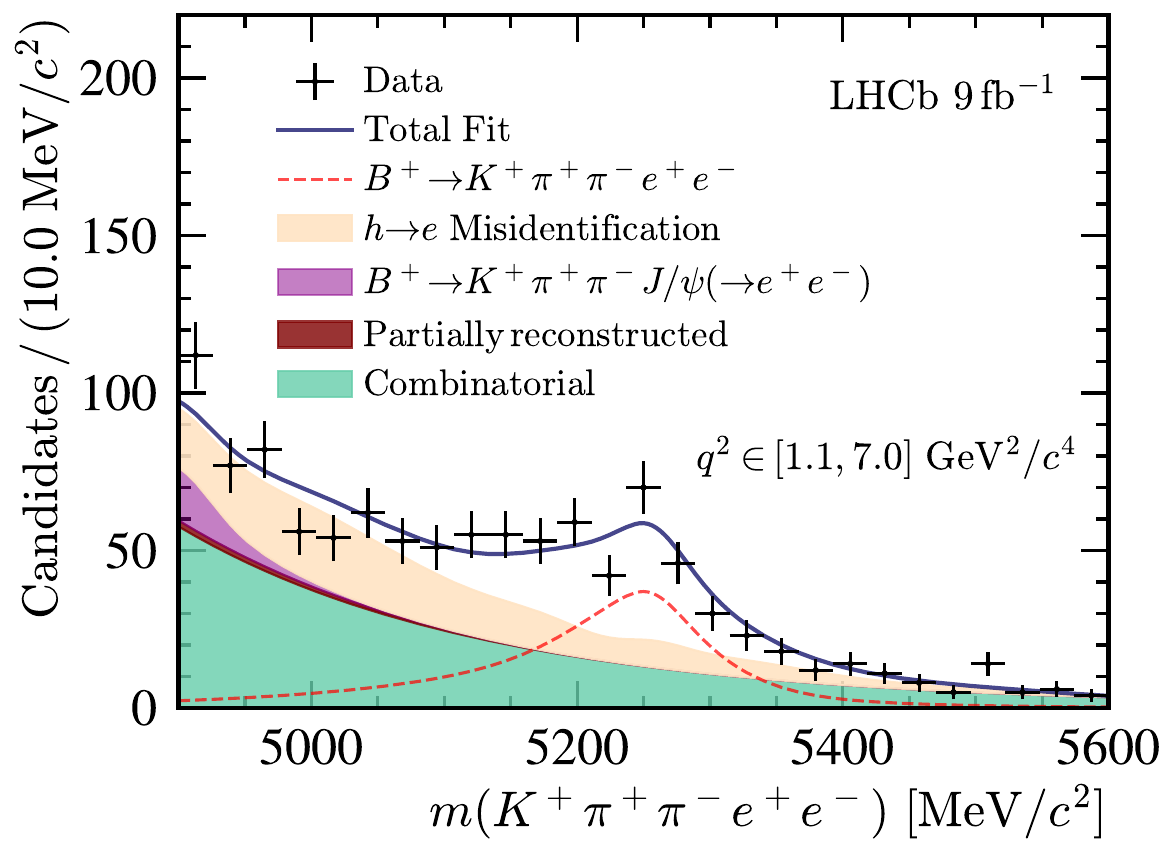}
     \includegraphics[width=0.45\linewidth]{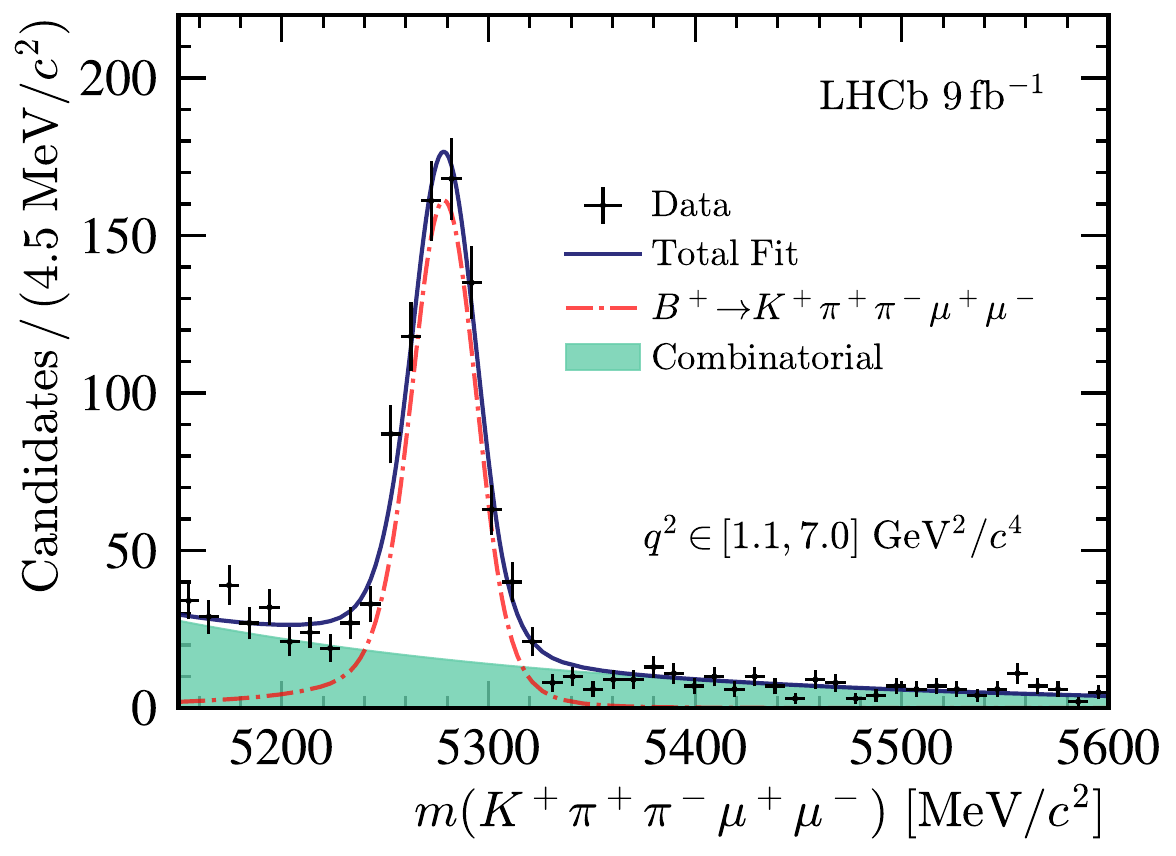} \\
     \includegraphics[width=0.45\linewidth]{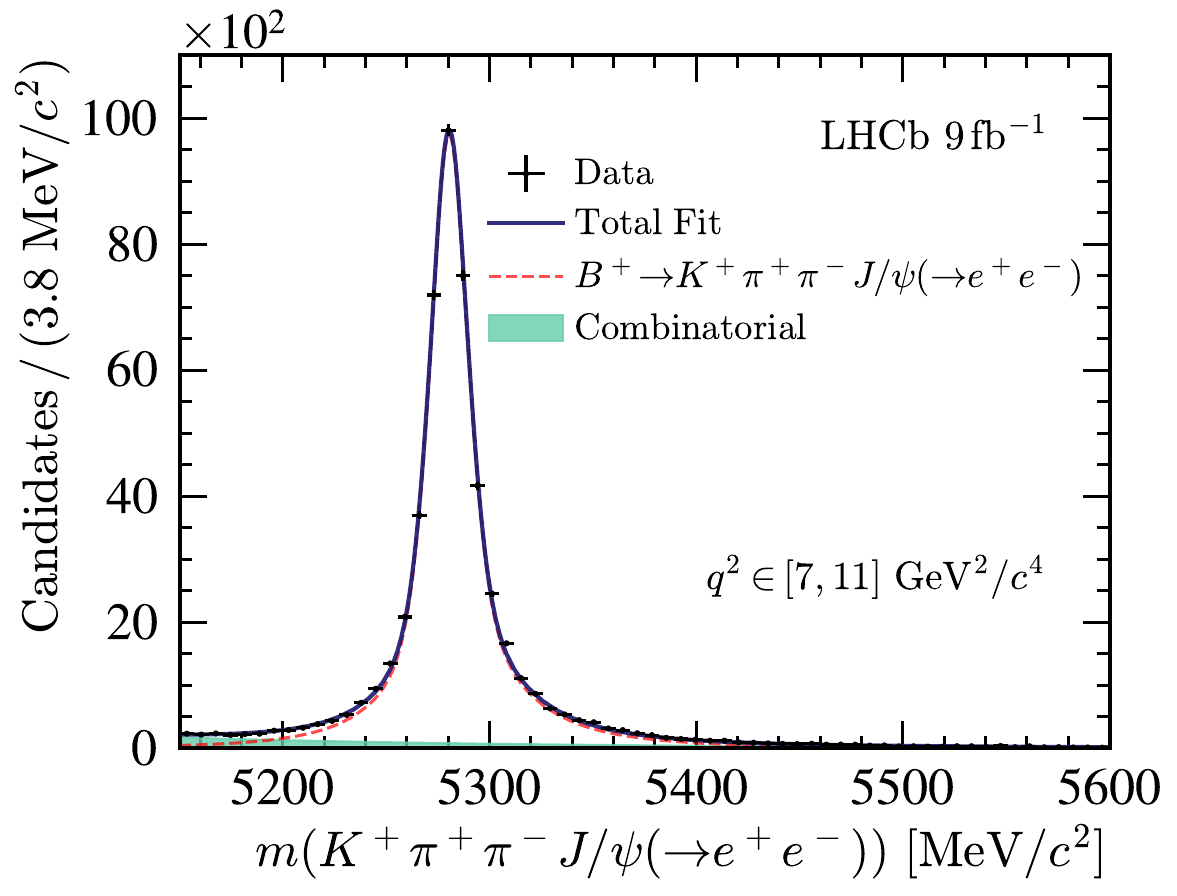}
     \includegraphics[width=0.45\linewidth]{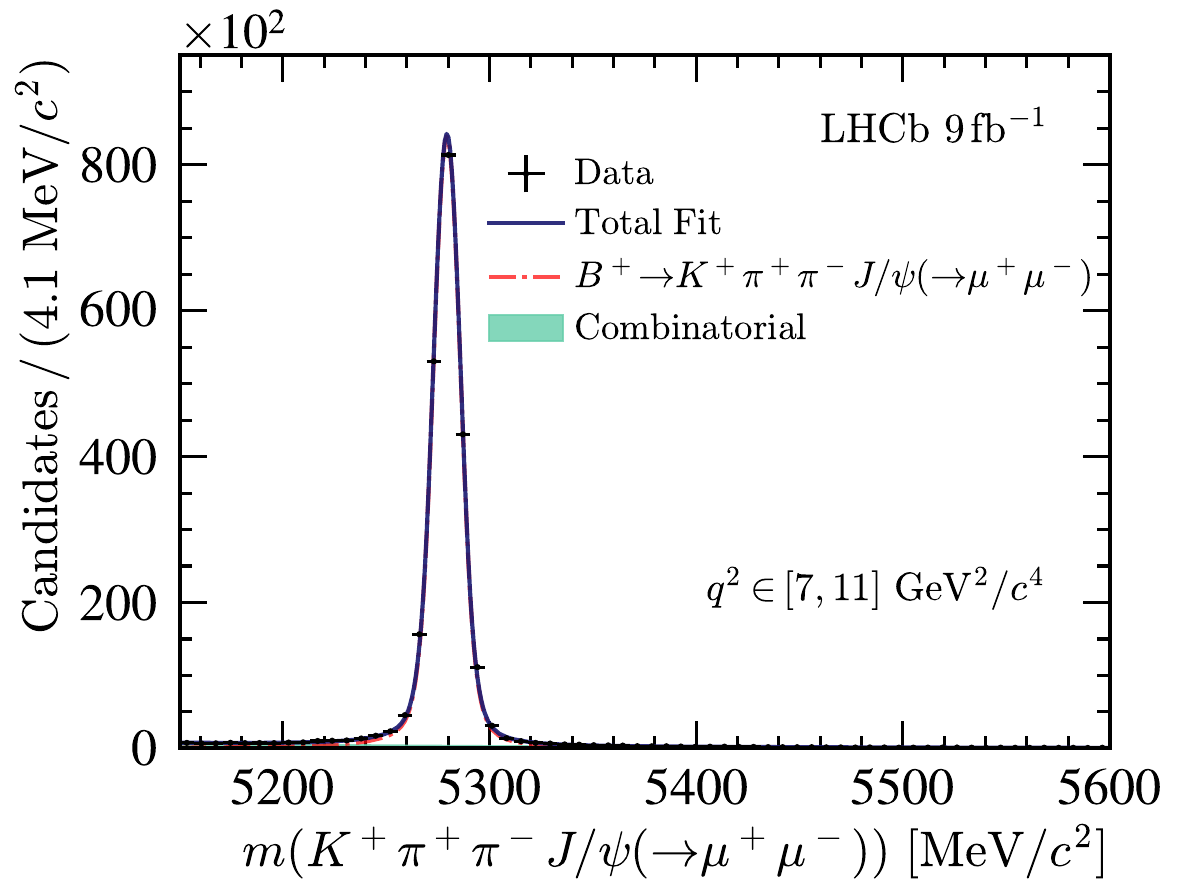} \\
   \end{center}
   \caption{
     Invariant-mass distributions together with the fit results of (top left) signal \mbox{\decayEE}, (top right) signal \mbox{\decayMM}, ~(bottom left) control \mbox{\decayJpsiEE} and ~(bottom right) control \mbox{\decayJpsiMM} decays, where the shaded background components are stacked.}
   \label{fig:fits}
 \end{figure}

The background due to hadrons misidentified as electrons is described following the procedure adopted by the \lhcb Collaboration for previous LFU tests~\cite{LHCb-PAPER-2022-045, LHCb-PAPER-2022-046}, with both its mass line shape and yield estimated directly from data. A sample of signal decays enriched in misidentified hadrons is obtained by inverting the PID requirements on the final-state electrons. The misidentified events are then weighted using misidentification rates measured in data from control channels, which allows an estimation of the amount of background candidates passing the signal selection criteria. A kernel-density estimator is used to model the mass of this background-enriched sample, whose line shape and yield are fixed in the fit to signal decays.

Results of the simultaneous fits performed to extract \rkpipiinv are shown in Fig.~\ref{fig:fits}. The measured signal yields are \mbox{$\mathcal{N}(\decayEE) = 264\pm21$} and \mbox{$\mathcal{N}(\decayMM) = 731\pm31$}, where the uncertainties only include the statistical component. The \rkpipiinv observable is measured to be
\begin{equation*}\label{eq:RKpipi_result}
   \rkpipiinv = 1.31^{+0.18}_{-0.17}\;(\mathrm{stat})\;^{+0.12}_{-0.09}\;(\mathrm{syst}),
\end{equation*}
where the first set of uncertainties are statistical and the second systematic. The result is in agreement with the SM expectation of LFU at the level of 1.7 standard deviations~($\sigma$).  The significance of the \mbox{\decayEE} decay is evaluated with Wilks' theorem~\cite{Wilks:1938dza} and found to exceed $10\sigma$, including both statistical and systematic uncertainties, confirming the first observation for the signal dielectron channel. The behavior of the likelihood profile is shown in Fig.~\ref{fig:NLL}, indicating the absence of other local minima.
\begin{table}[!t]
  \caption{Sources of relative systematic uncertainty considered in the measurement of \rkpipiinv. 
The contribution from each source is estimated by taking a weighted average of the systematic variations observed in each data subsample, split by data-taking period and hardware-trigger selection. To take into account correlations between subsamples, the values are scaled to the total systematic uncertainty obtained from the fit procedure, where the likelihood is convolved with a Gaussian distribution, with width equal to the quadratic sum of the systematic uncertainties for all sources considered. The total is evaluated by summing the different sources in quadrature.}\label{tab:syst}
  \begin{center}\begin{tabular}{lc}
  \hline
  Source & Uncertainty [$\%$] \\
  \hline
  \rjpsi nonflatness & $[-1.2, +1.6]$ \\
  Efficiency calibration & $[-1.8, +2.4]$\\
  Phase-space simulation & $[-3.0, +4.0]$ \\
  Fit bias & $[-1.1, 
  +1.4]$ \\
  Signal line shape & $[-1.7, +2.2]$ \\
  Leakage from resonant decays & $[-1.0, +1.4]$ \\
  Hadron-to-electron misidentification & $[-5.3, +7.1]$\\
  Partially reconstructed background & $[-0.9, +1.2]$ \\
  \hline
  Total & $[-6.9, +9.2]$\\
  \hline
\end{tabular}\end{center}
\end{table}

Table~\ref{tab:syst} summarizes all sources of systematic uncertainty considered. Systematic uncertainties associated with the efficiencies are evaluated for each correction applied to the simulated samples by using different input variables for the training of the reweighters or by using different decay modes as calibration samples. The correction of the phase-space model is the dominant systematic source related to the efficiencies. This systematic uncertainty is evaluated by training the reweighter on \mbox{\BuKpipiGamma} instead of \mbox{\decayJpsiMM} decays, assessing the effects of training the phase-space corrections at lower \qsq values as opposed to higher. Additional systematic uncertainties are assigned by training the reweighter on simulated signal samples where the \kpipi hadronic system is arbitrarily generated with several spin configurations.

The fit results are validated using an ensemble of pseudoexperiments generated from the nominal fit results. Subsequent fits of the pseudoexperiments are found to provide good statistical coverage, though with a small bias retained as a systematic uncertainty. 
Additional systematic uncertainties are assigned due to the modeling of components contributing to the invariant-mass spectrum, which might affect the estimation of signal yields. These are evaluated with the same pseudoexperiments used to assess the fit bias, refitting when varying the line shape or amount of a given fit component. The dominant systematic source of this type is related to the data-driven estimation of the background shape due the misidentification of hadrons as electrons. To determine this systematic uncertainty, the model used in the background-enriched sample is varied with an alternative empirical line shape consisting of the linear combination of two Gaussian functions with one exponential function. The uncertainty from the signal line shape is determined by fluctuating its shape parameters around the fixed values of the nominal fit. To assess the systematic uncertainty related to leakage from \mbox{\decayJpsiEE} decays into the signal sample, the yield of the leakage component is allowed to vary within its uncertainty obtained from simulation.
Finally, uncertainties induced by the partially reconstructed background model are evaluated using an alternative model that is derived fully from simulation.
Furthermore, in order to relax the Gaussian constraint relating the amounts of partially reconstructed background between signal and control decays, the Gaussian width is scaled by 150\% and the fit repeated, assigning the difference as a systematic uncertainty.

In summary, this Letter presents the first inclusive LFU test using \mbox{\decayLL} decays. Since the electron-mode measurement is statistically limited, further improvements are to be expected with the new dataset being collected with the \lhcb detector.
With $9\invfb$ of data, the measured value for \rkpipiinv is found to be compatible with the SM expectation of LFU at the $1.7\sigma$ level. These results provide independent constraints on possible LFU-violating extensions of the SM. 
The first observation of the \mbox{\decayEE} decay is also reported.

\section*{Acknowledgements}
%
% These Acknowledgements valid from 3-May-2019
%
\noindent We express our gratitude to our colleagues in the CERN
accelerator departments for the excellent performance of the LHC. We
thank the technical and administrative staff at the LHCb
institutes.
We acknowledge support from CERN and from the national agencies:
ARC (Australia);
CAPES, CNPq, FAPERJ and FINEP (Brazil); 
MOST and NSFC (China); 
CNRS/IN2P3 (France); 
BMBF, DFG and MPG (Germany); 
INFN (Italy); 
NWO (Netherlands); 
MNiSW and NCN (Poland); 
MCID/IFA (Romania); 
%MSHE (Russia); 
MICIU and AEI (Spain);
SNSF and SER (Switzerland); 
NASU (Ukraine); 
STFC (United Kingdom); 
DOE NP and NSF (USA).
%%%%%%%%%%%%%%%%%%%%%%%%%%%%%%%%%%%%%%%%%%%%%
We acknowledge the computing resources that are provided by ARDC (Australia), 
CBPF (Brazil),
CERN, 
IHEP and LZU (China),
IN2P3 (France), 
KIT and DESY (Germany), 
INFN (Italy), 
SURF (Netherlands),
Polish WLCG (Poland),
IFIN-HH (Romania), 
%RRCKI and Yandex LLC (Russia), 
PIC (Spain), CSCS (Switzerland), 
and GridPP (United Kingdom).
%%%%%%%%%%%%%%%%%%%%%%%%%%%%%%%%%%%%%%%%%%
We are indebted to the communities behind the multiple open-source
software packages on which we depend.
%%%%%%%%%%%%%%%%%%%%%%%%%%%%%%%%%%%%%%%%%%
Individual groups or members have received support from
%ARC and ARDC (Australia); % moved to national 16/01
Key Research Program of Frontier Sciences of CAS, CAS PIFI, CAS CCEPP, 
Fundamental Research Funds for the Central Universities,  and Sci.\ \& Tech.\ Program of Guangzhou (China);
Minciencias (Colombia);
EPLANET, Marie Sk\l{}odowska-Curie Actions, ERC and NextGenerationEU (European Union);
A*MIDEX, ANR, IPhU and Labex P2IO, and R\'{e}gion Auvergne-Rh\^{o}ne-Alpes (France);
%RFBR, RSF and Yandex LLC (Russia);
Alexander-von-Humboldt Foundation (Germany);
ICSC (Italy); 
%GVA, XuntaGal, GENCAT, Inditex, InTalent and Prog.~Atracci\'on Talento, CM (Spain);
Severo Ochoa and Mar\'ia de Maeztu Units of Excellence, GVA, XuntaGal, GENCAT, InTalent-Inditex and Prog.~Atracci\'on Talento CM (Spain);
SRC (Sweden);
the Leverhulme Trust, the Royal Society and UKRI (United Kingdom).

\setcounter{equation}{0}

\clearpage
\section*{End matter}

\begin{figure}[!htb]
  \begin{center}
    \includegraphics[width=0.65\textwidth]{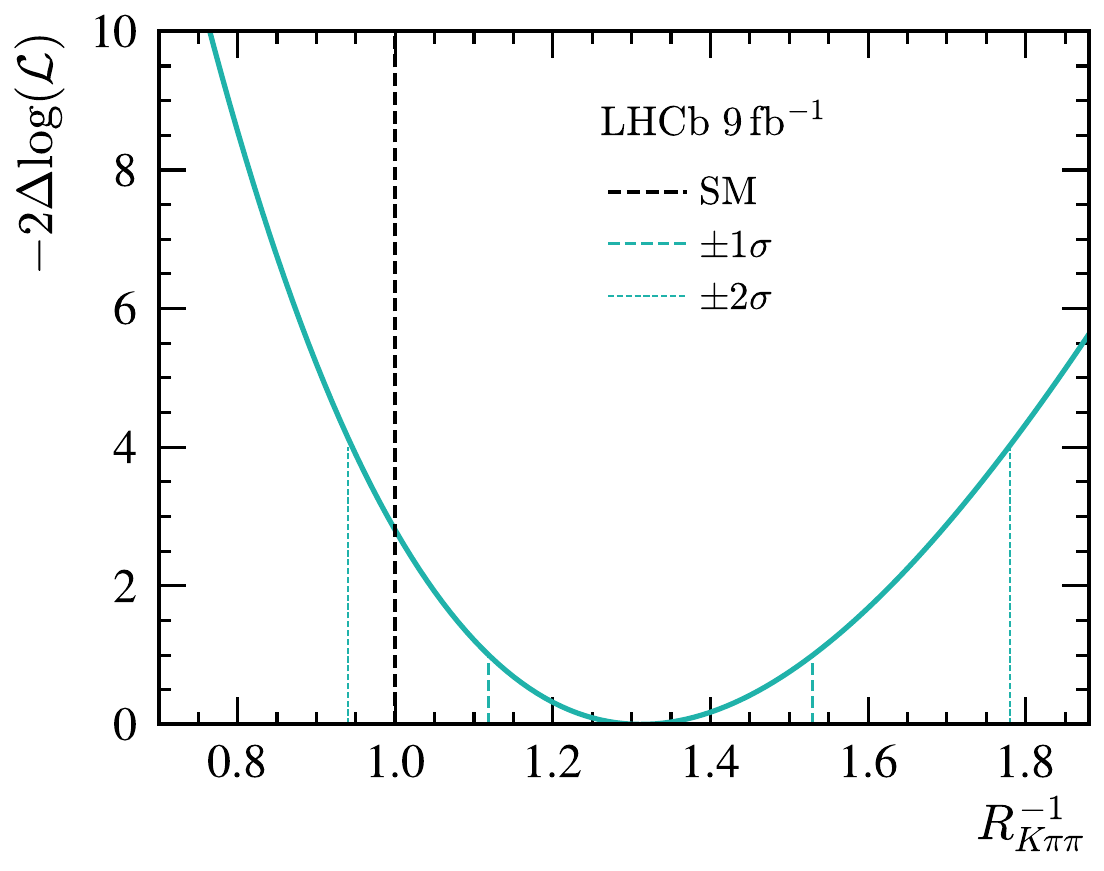}
    \vspace*{-0.5cm}
  \end{center}
  \caption{
    \small %captions should be a little bit smaller than main text
    Difference in twice the negative log-likelihood when varying \rkpipiinv with respect to the fitted value.}
  \label{fig:NLL}
\end{figure}

\figurename~{\ref{fig:NLL}} shows twice the negative log-likelihood profile scan from the maximum-likelihood fit used to extract \rkpipiinv. The likelihood profile includes systematic uncertainties, which are added directly in the fit procedure by convolving the likelihood with a Gaussian distribution, with width equal to the quadratic sum of the systematic uncertainties considered.

\clearpage

\addcontentsline{toc}{section}{References}
\bibliographystyle{LHCb}
\bibliography{main,standard,LHCb-PAPER,LHCb-CONF,LHCb-DP,LHCb-TDR}

\newpage
% LHCb collaboration author list
% Data extracted on December 11th, 2024 at 9:46am for paper reference LHCb-PAPER-2024-046
\centerline
{\large\bf LHCb collaboration}
\begin
{flushleft}
\small
R.~Aaij$^{38}$\lhcborcid{0000-0003-0533-1952},
A.S.W.~Abdelmotteleb$^{57}$\lhcborcid{0000-0001-7905-0542},
C.~Abellan~Beteta$^{51}$,
F.~Abudin{\'e}n$^{57}$\lhcborcid{0000-0002-6737-3528},
T.~Ackernley$^{61}$\lhcborcid{0000-0002-5951-3498},
A. A. ~Adefisoye$^{69}$\lhcborcid{0000-0003-2448-1550},
B.~Adeva$^{47}$\lhcborcid{0000-0001-9756-3712},
M.~Adinolfi$^{55}$\lhcborcid{0000-0002-1326-1264},
P.~Adlarson$^{82}$\lhcborcid{0000-0001-6280-3851},
C.~Agapopoulou$^{14}$\lhcborcid{0000-0002-2368-0147}
% ...

{\footnotesize \it

% ...
$^{14}$Universit{\'e} Paris-Saclay, CNRS/IN2P3, IJCLab, Orsay, France\\
% ...
$^{38}$Nikhef National Institute for Subatomic Physics, Amsterdam, Netherlands\\
% ...
$^{47}$Instituto Galego de F{\'\i}sica de Altas Enerx{\'\i}as (IGFAE), Universidade de Santiago de Compostela, Santiago de Compostela, Spain\\
% ...
$^{51}$Physik-Institut, Universit{\"a}t Z{\"u}rich, Z{\"u}rich, Switzerland\\
% ...
$^{55}$H.H. Wills Physics Laboratory, University of Bristol, Bristol, United Kingdom\\
% ...
$^{57}$Department of Physics, University of Warwick, Coventry, United Kingdom\\
% ...
$^{60}$School of Physics and Astronomy, University of Glasgow, Glasgow, United Kingdom\\
$^{61}$Oliver Lodge Laboratory, University of Liverpool, Liverpool, United Kingdom\\
% ...
$^{69}$Syracuse University, Syracuse, NY, United States\\
% ...
$^{82}$Department of Physics and Astronomy, Uppsala University, Uppsala, Sweden, associated to $^{60}$\\
% ...
}
\end{flushleft}

\end{document}